\documentclass[a4]{iopart}  
\usepackage[T1]{fontenc}
\usepackage[latin1]{inputenc}  
\usepackage{hyperref}  
\usepackage{geometry}
\usepackage{graphicx}  
\usepackage{amssymb}  
    
\date{\today}  
\def \be{\begin{equation}}
\def \bea{\begin{eqnarray}}
\def \eea{\end{eqnarray}}
\def \ee{\end{equation}}
\def \no{\nonumber}

\def \fsg{\frac{f}{g}}
\def \gsf{\frac{g}{f}} 
\def \a {\alpha}

\def \Ldot{\dot L}
\def \r{{\bf r}}
\def \eps{\epsilon}
\def \h{\frac{1}{2}}
\def\lsim{\mathrel{\rlap{\lower4pt\hbox{\hskip1pt$\sim$}}
    \raise1pt\hbox{$<$}}}                
\def\gsim{\mathrel{\rlap{\lower4pt\hbox{\hskip1pt$\sim$}}
    \raise1pt\hbox{$>$}}}                

\begin{document}  
\title{General relativistic treatment of LISA optical links}  
\author{S. V. Dhurandhar$^1$, J-Y. Vinet$^2$ and K. Rajesh Nayak$^3$}  
\address{ $^1$ IUCAA, Postbag 4, Ganeshkind, Pune - 411 007, India.   
\\ $^2$ ARTEMIS, Observatoire de la Cote d'Azur, BP 4229, 06304 Nice, France.  
\\$^3$ IISER-Kolkata, Block HC-7, Sector 3, Saltlake, Kolkata - 700106, India. 
}  
  
\begin{abstract}  
LISA is a joint space mission of the NASA and the ESA for detecting low frequency gravitational waves in the band $10^{-5} - 1$ Hz. In order to attain the requisite sensitivity for LISA, the laser frequency noise must be suppressed below the other secondary noises such as the optical path noise, acceleration noise etc. This is achieved by combining time-delayed data for which precise knowledge of time-delays is required. The gravitational field, mainly that of the Sun and the motion of LISA affect the time-delays and the optical links. Further, the effect of the gravitational field of the Earth on the orbits of spacecraft is included. This leads to additional flexing over and above that of the Sun. We have written a numerical code which computes the optical links, that is, the time-delays with great accuracy $\sim 10^{-2}$ metres - more than what is required for time delay interferometry (TDI) - for most of the orbit and with sufficient accuracy within $\sim 10$ metres for an integrated time window of about six days, when one of the arms tends to be  tangent to the orbit. Our analysis of the optical links is fully general relativistic and the numerical code takes into account effects such as the Sagnac, Shapiro delay, etc.. We show that with the deemed parameters in the design of LISA, there are symmetries inherent in the configuration of LISA and in the physics, which may be used effectively to suppress the residual laser noise in the modified first generation TDI. We demonstrate our results for some important TDI variables. 

\end{abstract} 

\pacs{95.55.Ym, 04.80.Nn, 07.60.Ly}

\maketitle  
\section{Introduction \label{SC:1}}  

A number of ground-based large-scale interferometric gravitational wave detectors, with 
optimal sensitivity in the frequency window $\sim$ 10 Hz -- 1 kHz are operational world-wide   \cite{GWD}. Among the large scale detectors, detectors of armlengths of few kilometres, the LIGO detectors of the US have now been continuously operating at initial design sensitivity which gives them a maximum range of about 25 Mpc for compact neutron star - neutron star binaries. The VIRGO detector of France and Italy has also attained comparable sensitivity. The detectors have been built with the {\it realistic goal of directly observing gravitational waves} (GW's) {\it for the first time}.
\par
A natural limit occurs on decreasing the lower frequency cut-off of $10$ Hz 
because it is not practical to increase the armlengths on ground and also because of 
the gravity gradient noise which is difficult to eliminate below $10$ Hz. 
The solution is to build an interferometer in space, where such noises will be absent and 
allow the detection of GW in the low frequency regime. LISA - Laser Interferometric Space 
Antenna - is a proposed mission which will use coherent laser beams exchanged between three 
identical spacecraft forming a giant (almost) equilateral triangle of side 
$5 \times 10^6$ kilometres to observe and detect low 
frequency cosmic GW \cite{RIP}. The ground-based detectors and LISA complement each other in the 
observation of GW in an essential way, analogous to the optical, radio, X-ray, 
$\gamma$-ray etc., observations do for the electromagnetic waves.
\par
In ground based detectors the arms are as symmetrical as possible so that the laser light 
experiences nearly identical delay in each arm of the interferometer. This arrangement reduces 
 the laser frequency/phase noise at the photodetector. Reduction of noise is 
crucial since the raw laser noise is orders of magnitude larger than other noises in the 
interferometer. But perfect symmetry is not possible, and an efficient system of 
servo loops is necessary for reaching a noise level compatible with the required sensitivity.
 The required sensitivity of the instrument can thus only be achieved by 
near exact cancellation of the laser frequency noise plus a good symmetry of the arms.
 However, in LISA, the lack of symmetry will be much larger than in terrestrial
 instruments, and the laser noise, though reduced by stabilisation techniques 
(still to be demonstrated), will probably be still too high. LISA consists  
of three correlated  interferometers, which produce redundancy in the data, and this can be used to suppress the laser frequency noise. In LISA, six data streams arise from the exchange of laser beams between the three spacecraft - it is not possible to bounce laser beams between different spacecraft, as is done in ground based detectors, because after 5 million km propagation,
the intensity of light reaching the target spacecraft is reduced by 10 orders of
magnitude; but in the target spacecraft a laser is locked in phase on the received wave,
so that the secondary beam is re-emitted without loss of phase information to the
primary source. This is analogous to the  RF transponder 
scheme, as was done in the early experiments for detecting GW by Doppler tracking 
a spacecraft from Earth \cite{Arm}. 
\par
Laser frequency noise which dominates the other noises by 7 or 8 orders of 
magnitude must be removed if LISA is to achieve the required sensitivity of $h \sim 10^{-22}$, where $h$ is the metric perturbation caused by a gravitational wave. This cancellation is achieved by time-delay interferometry (TDI) where the six data streams are combined with appropriate time-delays. This is possible because of the redundancy present in the data. This work was put on a sound footing by showing the data combinations had an algebraic structure; the data combinations cancelling laser frequency noise formed the {\it module of syzygies} over the polynomial ring of time-delay operators \cite{DNV}. This work was done for stationary LISA in flat spacetime where the motion of LISA as well as the ambient gravitation field, mainly that of the sun, was ignored. These were the so-called first generation TDI. However, LISA spacecraft execute a rotational motion and also the background spacetime is curved, all of which affect the optical links and the time-delays. Thus the Sagnac effect, Einstein effect, Shapiro delay, etc. are  important and must be incorporated into the analysis if the laser frequency noise is to be effectively cancelled. We here take into consideration all these effects in the full framework of general relativity. However, we compute the orbits of spacecraft in the  Newtonian framework. The base orbits we take to be Keplerian in the gravitational field of the Sun only, assuming the Sun to be a point mass. On these base orbits, we linearly superpose the perturbative effect of the Earth's gravitational field. We choose the Earth over Jupiter firstly because, the Earth perturbs the Keplerian orbit in resonance, resulting in unbounded growing of the perturbations and secondly, because of the technical reason that the Earth's effect can be easily accommodated within Clohessy-Wiltshire (CW)   framework \cite{CW}. Moreover we argue that Jupiter's effect is less than 10$\%$ of that of the Earth's and hence not a dominant one. The analytic approach helps to gain insight and understanding of the problem.
\par
Finally, given the arm flexing for our model of LISA, we compute the residual laser frequency noise spectrum for some important TDI observables, namely, the Sagnac, the Michelson and the Symmetric Sagnac in their modified first generation form. We find that the residual laser frequency noise in general tends not to be very high as compared with the secondary noises. If this level of noise is found to be acceptable, then there may be no need to use second generation TDI observables, which in general involve higher degree polynomials in time-delay operators and thus require more interpolations which in turn  result in larger errors in the data analysis. 
\par
We believe that the computations that we present here of the model would be of help in the development of a LISA simulator, the LISACode for instance \cite{LISACode}, because, (i) we have taken into account the relativisitic effects and (ii) the effect of the perturbation of the orbit due to the Earth. Also it would be useful to compare the model with actual data and look for any discrepancies. Any discrepancy arising could be interesting because that would imply the existence of some physical cause which has been overlooked and therefore would have to be incorporated into the data analysis.   
 
\section{The spacecraft orbits and flexing of LISA's arms}

We first describe the orbits in the gravitational field of the sun only. These are the usual Keplerian orbits. We then give the description of the same orbits in terms of the CW equations. This paves the way for including the effect of the Earth. Finally, we use the CW framework to include the perturbative effects of the Earth. 

\subsection{The Keplerian orbits of spacecraft in the Sun's field \label{SC:2A}}  

The Keplerian orbits are chosen so that the peak to peak variation in armlengths is the least $\sim 48000$ km, see \cite{NKDV}. We summarise the results of this paper below.    
We choose the Sun as the origin with Cartesian coordinates $\{X, Y, Z\}$ as follows:   
The ecliptic plane is the $X-Y$ plane and we consider a circular reference orbit   
of radius $R$ equal to 1 A. U. centred at the Sun.  Let $\delta_0 = 5 \a/8 $ where $\a = L_0/2R$  and $L_0 \sim 5,000,000$ km is a constant representing the nominal distance between two spacecraft of the LISA configuration. We choose the tilt of the plane of the LISA triangle to be $\delta = \pi/3 + \delta_0$ (this results in minimum flexing of the arms).  We choose spacecraft 1 to be at its lowest point (maximum negative Z) at $t = 0$.   This means that at this point, $Y = 0$ and $X \simeq R (1-e)$. The   
orbit of the first spacecraft is an ellipse with inclination angle   
$\epsilon_0$,  eccentricity $e$ and satisfying the above initial condition.  
\par
From the geometry, $\epsilon_0$ and $e$ are obtained as functions of $\delta$,  
\bea  
\tan \epsilon_0 &=&  \frac{\a \sin \delta}  
{\a \cos \delta + \sin(\pi/3)} \, ,\nonumber  \\   
e &=& \left[ 1+ \frac{4}{3} \a^2   + \frac{4}{\sqrt{3}}\a   
\cos \delta \right]^{1/2} - 1 \, .  
\label{eq:eincl}  
\eea  
The equations for the orbit of spacecraft 1 are given by:  
\bea  
X_1 &=& R(\cos \psi_1 - e) \cos \epsilon_0, \no \\  
Y_1 &=& R \sqrt{1 - e^2} \sin \psi_1, \no \\  
Z_1 &=& -R(\cos \psi_1 - e) \sin \epsilon_0.   
\label{tltorb}  
\eea    
The eccentric anomaly $\psi_1$ is implicitly given in terms of $t$ by,  
\be  
\psi_1 - e \sin \psi_1 = \Omega t - \phi_0\, ,  
\label{ecc}
\ee  
where $t$ is the time and $\Omega$ is the average angular velocity and $\phi_0$ the initial phase.
The orbits of the spacecraft 2 and 3 are obtained by rotating the orbit   
of spacecraft 1 by $2 \pi / 3$ and $4 \pi/3$ about the $Z-$axis; the phases $\psi_2, \psi_3$,    
however, must be adjusted so that the spacecraft are at a distance $\sim L_0$   
from each other. The orbital equations of spacecraft $k = 2, 3$ are:  
\bea  
X_k &=&  X_1 \cos \sigma_k- Y_1 \sin \sigma_k \, , \no \\  
Y_k &=&  X_1 \sin \sigma_k +  Y_1 \cos \sigma_k \, , \no \\  
Z_k &=& Z_1 \, ,  
\label{orbits}  
\eea  
where $\sigma_k = \left(k-1\right) \frac{2 \pi}{3}$,  
with the caveat that the $\psi_1$ is replaced by the phases $\psi_k$, where they   
are implicitly given by,  
\be  
\psi_k - e \sin \psi_k = \Omega t-\sigma_k - \phi_0.  
\label{eq:psik}  
\ee   
These are the exact (Keplerian) expressions for the orbits of the three spacecraft in the Sun's field.
In \cite{NKDV} it was shown that these orbits produce minimum flexing of LISA's arms when only the Sun's field is considered. Our next goal is to include the Earth's field and compute the flexing of LISA's arms in the combined field of the Sun and Earth. For this purpose we use the Clohessy-Wiltshire framework.

\subsection{The Clohessy-Wiltshire framework}

Clohessy and Wiltshire make a transformation to a frame - the CW frame $\{x, y, z\}$   
which has its origin on the reference orbit and also rotates with angular velocity $\Omega$.    
The $x$ direction is normal and coplanar with the reference orbit, the $y$ direction is  
tangential and comoving, and the $z$ direction is chosen orthogonal to the orbital plane.    
They write down the linearised dynamical equations for test-particles in the   
neighbourhood of a reference particle (such as the Earth). The length scale here is the Earth-Sun distance of 1 A. U. and the equations are applicable to distances small compared with this length scale. Since the frame is noninertial, Coriolis and centrifugal forces appear in addition to the tidal forces. The advantage of the CW equations is that it is easy to see that to the first order in $\a$ (or equivalently $e$) there exist configurations of spacecraft so that the mutual distances between them remain constant in time. The flexing appears only when we consider second and higher order terms in $\a$. In fact in \cite{NKDV} we find that the second order terms are sufficient to describe the flexing of LISA's arms quite accurately.     
\par
We take the reference particle to be orbiting in a circle of radius $R$ with constant angular velocity 
$\Omega$. Then the transformation to the  CW frame $\{ x, y, z \}$ from the barycentric frame $\{ X, Y, Z \}$ is given by,  
\begin{eqnarray}  
x & = & \left(X-R\,\cos\Omega t\right)\,\cos\Omega t\;+\;\left(Y-R\,\sin\Omega t\right)\,  
\sin\Omega t\,,\nonumber \\  
y & = & -\left(X-R\,\cos\Omega t\right)\,\sin\Omega t\;+\;\left(Y-R\,\sin\Omega t\right)\,  
\cos\Omega t\,,\nonumber \\  
z & = & Z.\label{eq:CW}  
\end{eqnarray}  
In the CW frame, the CW equations to second order in $\a$ (this includes upto the octupole field of the Sun) are,  
\bea  
\ddot{x}-2 \Omega \dot{y} - 3 \Omega^2 x + \frac{3 \a \Omega^2}{L_0}(2 x^2 - y^2 -z^2)& =& 0 \,  
 , \no \\  
\ddot{y} + 2 \Omega \dot{x} -\frac{6 \a \Omega^2}{L_0} x y& = &0 \, , \no\\  
\ddot{z} +  \Omega^2 z -\frac{6 \a \Omega^2}{L_0} x z& = &0.  
\label{gde2}  
\eea  
If we drop the terms in $\a$ in these equations we get the original CW equations (upto quadrupole). The solutions to these equations (that is upto quadrupole order) we call the zero'th order. Among these we choose the solutions which form an equilateral triangular configuration of side $L_0$. For the $k$th spacecraft we have the  following coordinates:
\bea
x_k &=&-\frac{1}{2} \rho_0 \cos(\Omega t - \sigma_k - \phi_0) \, , \no \\
y_k &=& \rho_0 \sin( \Omega t - \sigma_k - \phi_0) \, , \no \\
z_k &=& -\frac{\sqrt{3}}{2}\rho_0 \cos(\Omega t - \sigma_k - \phi_0) \, ,
\label{cws}
\eea 
where $\rho_0 = L_0 /\sqrt{3}$. Also at $t=0$ the initial phase of the configuration is described through $\phi_0$. In this solution, any pair of spacecraft maintain the constant distance $L_0$ between each other.
\par
In \cite{NKDV} we have shown that if include the $\a$ terms (octupolar terms) and solve perturbatively using the zeroth order solution as given by Eq.(\ref{cws}), we obtain the flexing of the arms. Further, this approximate solution agrees to a remarkable degree with the flexing deduced from the exact Keplerian orbits.  

\subsection{The effect of the Earth}

LISA follows the Earth $20^{\circ}$ behind. We consider the model where the centre of the Earth leads the origin of the CW frame by $20^{\circ}$ - thus in our model, the `Earth' or the centre of force representing the Earth, follows the circular reference orbit of radius 1 A. U. Also the Earth is at a fixed position vector $\r_{\oplus} = (x_{\oplus}, y_{\oplus}, z_{\oplus})$ in the CW frame. We find that $x_{\oplus} = - R (1 - \cos 20^{\circ}) \sim - 9 \times 10^6$ km, $y_{\oplus} = R \sin 20^{\circ} \sim 5.13 \times 10^7$ km and $z_{\oplus} = 0$, where we have taken $R$ to be 1 A. U.. The force field ${\bf F}$ due to the Earth at any point $\r$ (in particular at any spacecraft) in the CW frame is given by:
\be
{\bf F} (\r) = - G M_{\oplus} \frac{\r - \r_{\oplus}}{|\r - \r_{\oplus}|^3} \,,
\ee   
where $M_{\oplus} \sim 5.97 \times 10^{24}$ kg is the mass of the Earth and 
$G = 6.67 \times 10^{-11} ~ {\rm kg}^{-1} {\rm m}^3 {\rm sec}^{-2}$ Newton's gravitational constant. 
\par
In order to write the CW equations in a convenient form we first define the small parameter $\eps$ in terms of the quantity $\omega_{\oplus}^2 = G M_{\oplus} / d_{\oplus}^3$, where 
$d_{\oplus} = |\r_{\oplus}|$ is the distance of the Earth from the origin of the CW frame; $d_{\oplus} \sim 5.2 \times 10^7$ km which is more than 50 million km. So when deriving the forcing term we make the aprroximation $|\r - \r_{\oplus}| \approx d_{\oplus}$, that is, we neglect $|\r|$ compared to 
$d_{\oplus}$. It will turn out that the flexing due to the Earth is small so that this approximation is not unjustified. We define $\eps = \omega_{\oplus}^2 / \Omega^2 \simeq 7.16 \times 10^{-5}$ which is the just the ratio of the tidal forces due to the Earth and the Sun. The CW equations including the Earth's field take the form:
\bea  
\ddot{x}-2 \Omega \dot{y} - 3 \Omega^2 x + \eps \Omega^2 (x - x_{\oplus})& =& 0 \,, \no \\  
\ddot{y} + 2 \Omega \dot{x} + \eps \Omega^2 (y - y_{\oplus}) & = & 0 \,, \no\\  
\ddot{z} +  \Omega^2 (1 + \eps) z & = &0.  
\label{CWearth}  
\eea
Note that the compounded flexing due to the combined field of Earth and Sun is nonlinear; it is infact a three body problem. We however solve this problem approximately. Assuming that both effects are small we may linearly add the flexing vectors due to the Sun and Earth; that is, add the perturbative solutions obtained from Eqs.(\ref{gde2}) and (\ref{CWearth}); the nonlinearities appear at higher orders in $\a$ and $\eps$. These would modify the flexing but we may neglect this effect because of the smallness. As it will turn out, the flexing produced due to the Earth is of the order of 1 or 2 m/sec upto the third year, just about 40 $\%$ of that due to the Sun. But, as shown in \cite{NKDV} the flexing produced by the Sun's octupole field is nearly exact to that produced by the Keplerian orbits. Thus we may do better by linearly adding the flexing vector produced by the Earth to the Keplerian orbit of the relevant spacecraft. We therefore, first compute the motion of spacecraft perturbatively using the zero'th order solutions as given in Eq.(\ref{cws}). This will induce a flexing of the LISA arms only by the Earth's field.  
\par   
We now seek perturbative solutions to Eq. (\ref{CWearth}) to the first order in $\eps$. We write, 
$x = x_0 + \eps x_1, y = y_0 + \eps y_1, z = z_0 + \eps z_1$ where $x_0, y_0, z_0$ are solutions at the zeroth order given by Eq.(\ref{cws}). We put $\sigma_k = 0$ (or equivalently include it in $\phi_0$) in these solutions for simplifying the algebra. 
\par
Note that the $z$ equation is decoupled from the $x$ and $y$ equations which are themselves coupled; infact the $z$ equation can be solved exactly.  We also assume the initial conditions for the perturbative solutions to be homogeneous, that is, we take, $x_1 = y_1 = z_1 = \dot{x}_1 = \dot{y}_1 = \dot{z}_1 = 0$ at $t = 0$ - the spacecraft are in the desired positions initially. 
\par
We first solve the $x$ and $y$ equations. To the first order in $\eps$, the equations for these  perturbations are:
\bea
\ddot{x}_1-2 \Omega \dot{y}_1 - 3 \Omega^2 x_1 &=& \Omega^2 x_{\oplus} + \h \Omega^2 \rho_0 \cos (\Omega t - \phi_0) \,, \no \\  
\ddot{y}_1 + 2 \Omega \dot{x}_1 &=& \Omega^2 y_{\oplus} - \Omega^2 \rho_0 \sin (\Omega t - \phi_0)  \,.
\label{prtxy}  
\eea 
We note that the forcing terms on the right hand side of these equations appear at the same frequency $\Omega$ and hence they imply resonance. This means that the Earth's effect on LISA is cumulative as the detailed calculations show below. Therefore, it is most important to include the effect of the Earth on LISA.
\par
The equation for $y_1$ can be easily integrated, once, with the initial conditions mentioned above and the $\dot{y}_1$ substituted in the $x_1$ equation resulting in a decoupled equation for $x_1$. This decoupled equation can be solved to yield the solution for the perturbation $x_1$. The solution $x_1$ in turn can be substituted back into the $y_1$ equation to obtain a first order equation for $y_1$ and integrated with the initial conditions. Without further ado we state the results:
\bea
x_1 &=& - \rho_{\oplus} \cos (\Omega t - \phi_{\oplus}) + x_{\oplus} + 2 y_{\oplus} \Omega t - 2 \rho_0 \cos \phi_0 + \frac{5}{4} \rho_0 \Omega t \sin (\Omega t - \phi_0) \,, \no \\
y_1 &=& 2 \rho_{\oplus} [\sin (\Omega t - \phi_{\oplus}) + \sin \phi_{\oplus}] - \frac{3}{2} \rho_0 [\sin (\Omega t - \phi_0) + \sin \phi_0] \no \\
&& + \frac{5}{2} \rho_0 \Omega t \cos (\Omega t - \phi_0) - \Omega t (2 x_{\oplus} - 3 \rho_0 \cos \phi_0) - \frac{3}{2} \Omega^2 t^2 y_{\oplus} \,,
\label{solnxy}
\eea
where,
\bea
\rho_{\oplus}^2 &=& (x_{\oplus} - 2 \rho_0 \cos \phi_0)^2 + (2 y_{\oplus} - \frac{5}{4} \rho_0 \sin \phi_0)^2  \,, \no \\
\tan \phi_{\oplus} &=& \frac{2 y_{\oplus} - \frac{5}{4} \rho_0 \sin \phi_0}{x_{\oplus} - 2 \rho_0 \cos \phi_0} \,.
\eea
The $z$ equation can be exactly integrated and used directly to obtain the flexing. However, we can also expand this solution to the first order in $\eps$ and the result is:
\be
z_1 = \frac{\sqrt{3}}{4} \rho_0 [\Omega t \sin \Omega t \cos \phi_0 - (\Omega t \cos \Omega t - \sin \Omega t) \sin \phi_0] \,.
\label{solnz}
\ee
We observe that since $y_{\oplus}$ is very large compared to other distances in $\rho_{\oplus}$, we have $\rho_{\oplus} \sim 2 y_{\oplus} \sim 10^8$ km. Secondly $\phi_{\oplus} \sim \pi / 2$ for the same reason. For spacecraft 1, for instance, with the initial condition $\phi_0 = 0, \phi_{\oplus} \sim 91.83^{\circ}$. This aids in simplifying much of the computations.  
\par
We now turn to the flexing of the arms for which we must compute the perturbation of each spacecraft orbit due to the Earth. As argued previously, we merely add the perturbation given by 
$\eps \r_1 = \eps (x_1, y_1, z_1)$ to the Keplerian orbit of each spacecraft. For spacecraft 1, the Keplerian orbit is given by  Eq. (\ref{tltorb}) and Eq. (\ref{ecc}) in the barycentric frame. We denote this zeroth order orbit by the vector trajectory ${\bf R}_0 (t; \phi_0)$ having the initial phase $\phi_0$. We then choose the zeroth order solution in the CW frame corresponding to this orbit which is given by Eq. (\ref{cws}) with $k=1$ or $\sigma_1 = 0$. With this solution we compute the perturbative solution $\r_1 (t; \phi_0)$ and finally obtain the total orbit ${\bf R}_1 (t; \phi_0) = {\bf R}_0 (t; \phi_0) + \eps \r_1 (t; \phi_0)$. We repeat the same procedure for the other two spacecraft and obtain the orbits which now include the effect of the Earth as well. The flexing of the LISA arms is now due to both the Sun and the Earth. Below in Figure \ref{earth_flex}, we plot the flexing of LISA's arms as a function of $t$, assuming constant length $L_0$ in the Sun's field, that is, we take ${\bf R}_0$ as given from Eq.(\ref{cws}) so in effect we are considering only the flexing due to the Earth. We also take $\phi_0 = 0$ as the initial condition.
\begin{figure}[h!]  
\centering  
\includegraphics[width = 0.6\textwidth]{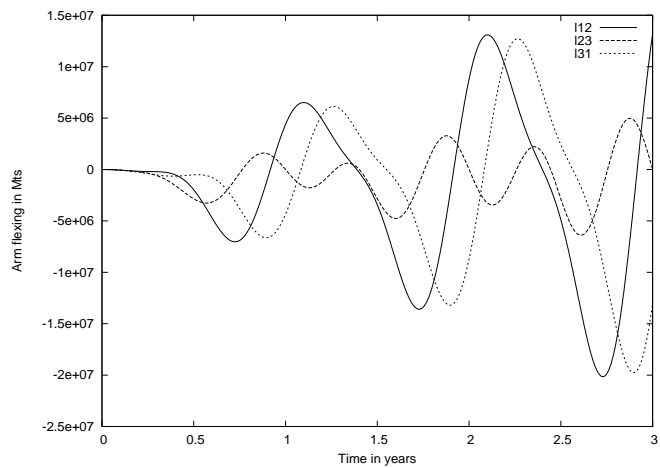}  
\caption{The figure shows the effect of only the Earth on flexing of the three arms of LISA for a period of three years. The flexing grows to a maximum of about 20,000 km in the third year which is about 40$\%$ of the flexing due to the Sun.}  
\label{earth_flex}  
\end{figure}
\par
We notice that the flexing increases with time; in third year it grows to a maximum of about 20,000 km from the initial value. This variation is about 40$\%$ of that due to the Sun - $\sim 48,000$ km. Of course, one must consider the vector additions, so that the total flexing will in general be less than this or even reduce if the flexing vectors are oppositely oriented. Also by looking at the slope of the curves we note that the rate of change of armlength is of the order of a metre/sec in the the second year and less than 2 metres/sec in the third year. Thus at least in the first three years we do not expect the flexing to be affected too much by the Earth. The flexing also depends on the initial phase $\phi_0$, that is, the epoch at which the gravitational field of the Earth is `switched on'. Here there is also a symmetry; the flexing profiles remain invariant if we change $\phi_0$ to $\phi_0 + \pi/3$. Although, for different $\phi_0$ the flexing profiles differ in detail, they display the similar qualitative behaviour.  
\par
The next task is to compute the optical links with the above spacecraft orbits from which the total flexing due to the Sun and Earth can be deduced. 

\section{The numerical computation of optical links}

The time-delay that is required for the TDI operators needs to be known very accurately - at least to 1 part in $10^8$, that is, to about few metres - for the 
laser frequency noise to be suppressed. In order to guarantee such level of accuracy, we have to  numerically compute the optical links or the time-delay. This approach is guaranteed to give the desired accuracy or even better accuracy than what is required. In this approach, we numerically integrate the null geodesics followed by the laser ray  emitted by one spacecraft and received by the other. This computation is performed in the barycentric frame, and taking into account the fact that the spacetime is curved by the Sun's mass. The computation here is further complicated by the fact that the spacecraft are moving in this frame of reference and the photon emitted from one spacecraft must be received by the other spacecraft. We use the Runga-Kutta numerical scheme to integrate the differential equations describing the null geodesics. But since the end point of the photon trajectory is not known apriori, an iterative scheme must be devised for adjusting the parameters of the null geodesic, in order that the worldlines of the  photon and the receiving spacecraft intersect. We have devised such a scheme based on the difference vector between the photon position vector and receiving spacecraft position vector. The six optical links $L_{ij}$ have thus been numerically computed with sufficient accuracy required for TDI: for 98$\%$ of the time the code we have devised gives excellent results to the accuracy of $10^{-2}$ metres with $10^5$ steps. For the rest of the time, $2 \%$, when the any one of the arms tends to lie tangent to the orbit, one must increase the number of steps. We increase the number of steps to $10^7$. Then the code gives results accurate upto 10 metres except in a window of about half an hour when the error exceeds this value and becomes unacceptably large. This is because the differential equations describing the null geodesics encounter sign changes in the components of the tangent vector which must be carefully incorporated into the integration scheme. Such windows occur six times in a year two months apart. In this paper we choose to ignore these time windows.

\subsection{Optical links: integrating the null geodesic equations \label{SC:2B}}

We now turn to the optical links. The spacetime geometry taking only the Sun as the gravitating mass is given by the Schwarzschild spacetime whose metric in isotropic coordinates is described by:
\be
ds^2 = f(r)c^2 dt^2 - g(r)[dr^2 + r^2 (d \theta^2 + \sin^2 \theta d \phi^2)] \,,
\ee
where the functions $f(r)$ and $g(r)$ are given by:
\be
f(r) = 1 - \frac{2m}{r}, ~~~~g(r) = 1 + \frac{2m}{r} \,.
\ee 
Here $m = GM/c^2$ where $M$ is the mass of the Sun, $c$ the speed of light and $G$ the Newton's gravitational constant.
\par 
The null geodesics satisfy the differential equations:
\bea
{\dot r} &=& {\epsilon_r} \fsg \sqrt{\gsf - \frac{b^2}{r^2}}, \\
{\dot \theta} &=& {\epsilon_{\theta}} \frac{1}{r^2} \fsg \sqrt{b^2 - \frac{L^2}{\sin^2 \theta}}, \\
{\dot \phi} &=& \fsg \frac{L}{r^2 \sin^2 \theta} \,.
\eea
Here $\epsilon_{r,\theta}$ are the appropriate signs for the orbit equations, the overdot is $d /c dt$, where $c$ is the speed of light, $b$ is the impact parameter and $L$ the azimuthal angular momentum. 
\vspace{0.2in}

Since the spacecraft are moving, the parameters $b$ and $L$ are not known apriori, but should be obtained from the solution by an iterative scheme. Let us first consider the optical link, say link 12, from S/C 1 (position vector ${\bf r_1}(t)$) to S/C 2 (position vector ${\bf r_2}(t)$) where $t$ is the coordinate time in the isotropic coordinates. Since the space is almost flat, we take the zero-th order estimate as flat spacetime. Then the initial estimate of $b$ is given by:
\be
b_{(0)} = |{\bf r_1(t_0)} \times {\bf n}| 
\ee
where $t_0$ is the time at which S/C 1 emits the photon. Similarly for $L$, the zero-th order estimate denoted by $L_{(0)}$ is the z-component of the vector ${\bf r_1}(t_0) \times {\bf n}_{(0)}$ and ${\bf n}_{(0)}$ points towards S/C 2. To start with a better estimate of ${\bf n}_{(0)}$ we take the better estimate of the delayed position of S/C 2 at $t + L_0/c$ (note this is still not the correct final position of S/C 2, when the photon meets S/C 2).  The null geodesic (photon orbit) is then integrated till the time as required in flat spacetime. The photon obviously does not hit the S/C 2, because (i) S/C 2 has moved in the meanwhile, (ii) Shapiro delay: the photon is delayed because of the gravitational field of the Sun. The difference vector between the photon and S/C 2 drives the iterative scheme: we decompose this vector into parallel and perpendicular components with respect to ${\bf n}_{(0)}$ and use these projections to obtain a modified direction say ${\bf n}_{(1)}$ (perpendicular component) and a new time of flight (parallel component). The ${\bf n}_{(1)}$ produces new values of $b$ and $L$, say, $b_{(1)}, L_{(1)}$ and so on until convergence is reached to the desired accuracy. 
\par

We divide our discussion into two parts: (i) normal time epochs and (ii) anomalous time epochs $(b \sim L)$. The anomalous time epochs, occur sixth of an year apart when any one of the arms lies tangential to the orbit, that is, for the relevant null geodesic $b \sim L$. With the initial conditions, we have chosen, the first such epoch for the link 12 occurs at 
$\sim 52 \times 10^5$ secs. In the normal case, we use $10^5$ steps in the Runga-Kutta scheme, each step of about 50 km. Just 2 or 3 iterations suffice to produce the necessary convergence and with an accuracy to $10^{-2}$ metres. In the anomalous case, we need more steps in the Runga-Kutta scheme; we use $10^7$ steps which gives an accuracy within 10 metres except for a window of about half hour. In the anomalous case at least one of the ${\dot \theta}$ and ${\dot r}$ (or equivalently $\eps_r$ and $\eps_{\theta}$) change sign along the null geodesic and this sign change must be taken into account in the Runga-Kutta integration scheme as close as possible to the turning point. We achieve this by reducing the step size to $\sim 0.5$ km, so that the error remains at an acceptable level. Even then in a window of about half an hour, every six months, the error becomes large. We ignore these points by smoothly interpolating.  
\par
\begin{figure}[h!]  
\centering  
\includegraphics[width = 0.6\textwidth]{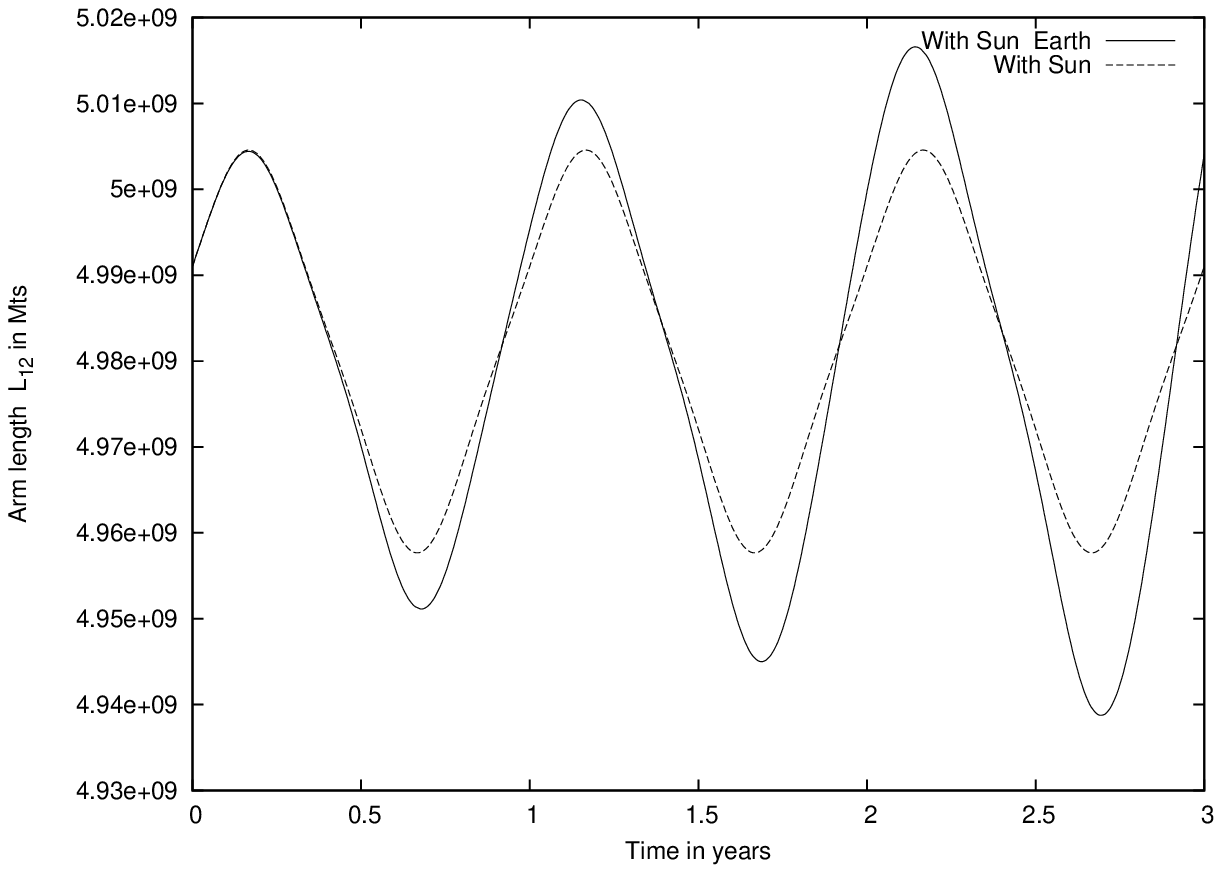}  
\caption{Comparison of flexing just due to Sun (dotted curve) and the combined effect of Sun and Earth (bold curve). The flexing of the arm is more and especially high in the third year in the combined field of the Sun and Earth as compared to that of Sun only. The figure is plotted for a duration of three years and with $\phi_0 = 0$ for the link 12.}  
\label{cmpr}
\includegraphics[width = 0.6\textwidth]{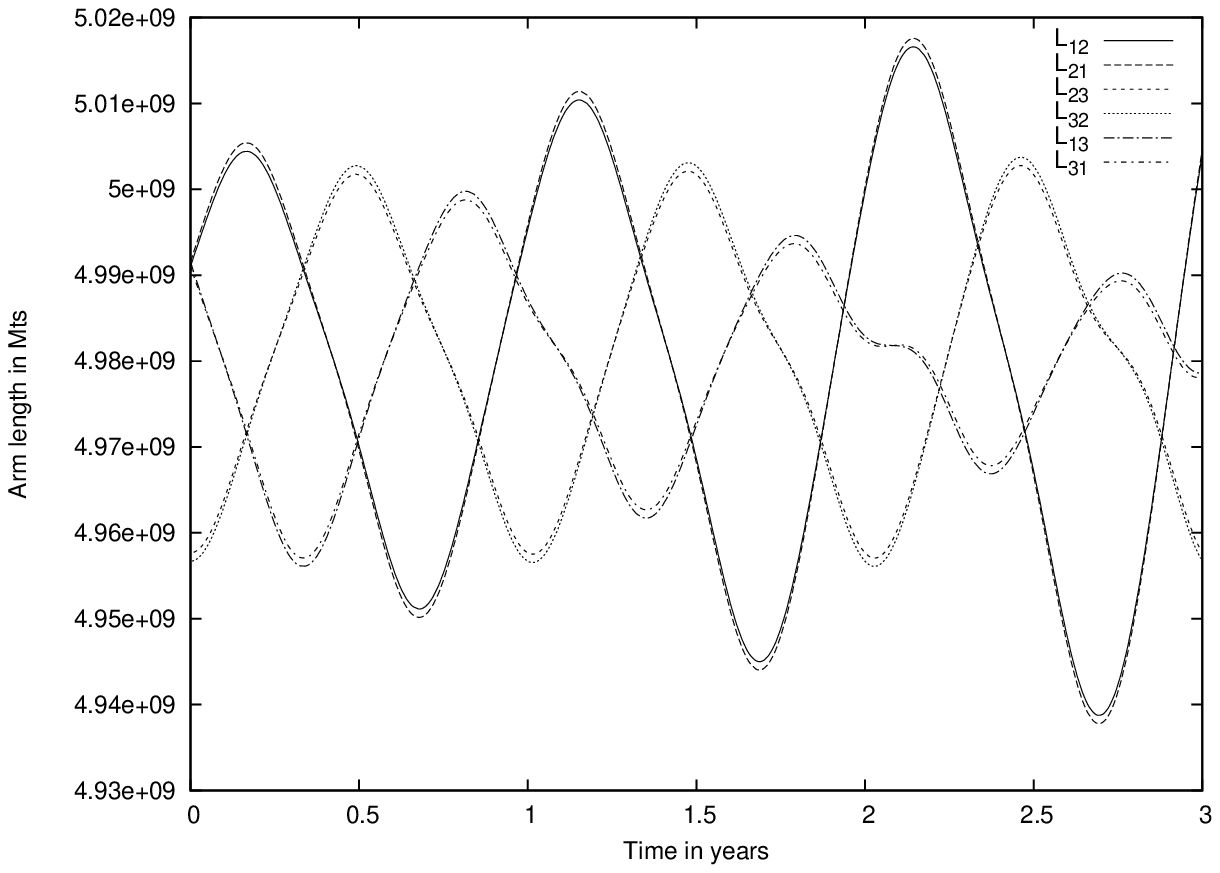}  
\caption{The figure shows the variation in the six optical links of the LISA model for three  years. The lengths are given in metres.}  
\label{optlink}  
\end{figure}

The Figure \ref{cmpr} shows the comparison of the optical link $L_{12}$ just due to the Sun and when the Earth's field is also included. The flexing increases by $\sim 20,000$ km (as compared to $L_0$) when the Earth's field is included. The figure is drawn for a time period of about three years and $\phi_0$ is chosen to be zero. The important difference due to the combined field is that the motion of spacecraft and hence the variations in armlengths is no more periodic and infact grows with time. For certain links (in our case links 23 and 32), the flexing is infact reduced because the flexing vectors due to the Sun and Earth are oppositely oriented. 
\par
Figure \ref{optlink} shows all the six optical links in the combined field of the Sun and Earth. 

\subsection{Flexing of the arms \label{SC:2C}}
\vspace{0.2in}

\begin{figure}[h!]  
\centering  
\includegraphics[width = 0.6\textwidth]{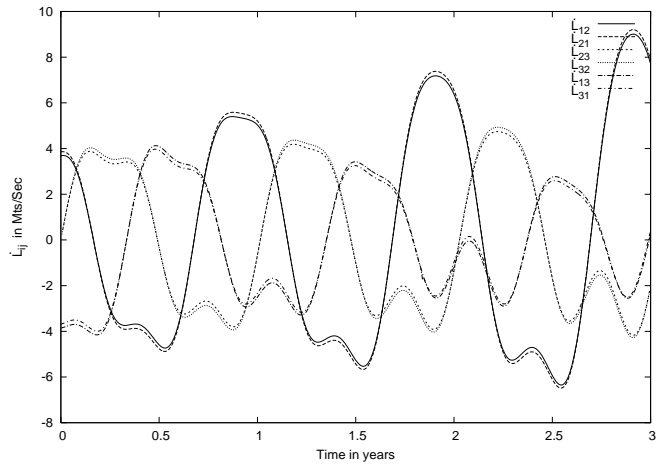}  
\caption{The rate of change of armlengths for the six links is shown in units of m/sec. This rate of change is less than 6 m/sec upto the second year and increases to a maximum of about 
8 m/sec in the third year.}  
\label{dplr}  
\end{figure}

We also need to estimate the variation in armlength which is important for the TDI analysis to follow. Although there has been some {\it adhoc} work on this topic previously, the question needs to be revisited and a complete solution sought. Here we propose to address the question of the residual noise, having given the exact optical links. The first important task is to exactly compute the rate of change of arm-length. Figure \ref{dplr} shows the rate of change of the six optical links as a function of time over a period of three years.

\par
We find that in the optimised model of LISA configuration, this rate of change is less than 4 m/sec. if we just consider the Sun's field. Including the Earth's field the flexing still remains $\lsim 6$ m/sec in the first two years and increases to $\lsim 8$ m/sec in the third year. Earlier estimates were $\sim 10$ m/sec. This numerical estimates are most crucial for their effect on residual laser frequency noise in the TDI. 

\section{Time-delay interferometry with variable armlengths}

 In order to cancel the laser frequency noise, time-delayed data streams are added together in which an appropriate set of time-delays are chosen. In general the time-delays are multiples of the photon transit time between pairs of spacecraft. In \cite{DNV} a scheme based on modules over commutative rings was given where the module of data combinations cancelling the laser noise was constructed. This fully cancels the laser frequency noise for stationary LISA. There are only three delay operators corresponding to the three armlengths and the time-delay operators commute. This scheme can be straight forwardly extended to moving LISA \cite{NV}, where, now because of Sagnac effect, the up and down optical links have different armlengths (photon transit time) but the  armlengths are still  constant in time. Now there are six delay operators corresponding to the six optical links and they commute. These are the modified but still first generation TDI.  It is crucial for the operators to commute if this scheme is to work - we must have a commutative ring. However, for LISA the armlengths do change as a function of time - flexing of the arms - and the first generation TDI modified or otherwise lead to imperfect cancellation of the laser frequency noise. Here we consider the model for LISA whose variation in arm-length is minimum when only the Sun's field is considered. This property may no more exactly hold in the combined field of the Sun and Earth, but assuming that Sun's effect is dominant the variation in armlength may still continue to be near optimal. Here, we however, assume the same initial conditions and compute the residual laser noise in the first generation modified TDI variables. 
\par
Let $C(t) = \Delta \nu(t)/\nu_0$ represent the laser frequency noise in some optical link. Let $D_j$ be the delay operator corresponding to the armlength $L_j (t)$,i.e. $D_j C(t) = C(t - L_j (t))$. If we operate on $C(t)$ with operators $D_j$ and $D_k$ in different orders, it is easily seen that 
\be
D_j D_k C (t) \neq D_k D_j C(t).
\ee
The operators do not commute. A combinatorial approach has been adopted in \cite{Vallis} to deal with the totally noncommutative case. However, our aim here is to estimate the level of the noncommutativity of these operators in the context of the LISA model, compute the residual laser frequency noise and compare it with the other secondary noises in LISA for several of the TDI combinations. Our investigations will show whether the noncommutativity can be ignored or one must deal with non-commutative operators. Here our investigations necessarily take into account the symmetry of the LISA configuration and therefore  we expect the residual laser frequency noise to be smaller than if the symmetries were absent. 
\par
The first step is to develop a calculus of the $D_j$ operators. 

\subsection{The calculus of the delay operators and relevant commutators}

We find that for this calculation we require to develop this only to the first order in ${\dot L}$. This is because we find for our model $\ddot L \sim 10^{-6}$ metres/sec$^2$ and thus even if one considers say 6 successive optical paths, that is, about $\Delta t \sim 100$ seconds of light travel time, $\Delta t^2 {\ddot L} \sim 10^{-2}$ metres. This is well below few metres and thus can be neglected in the residual laser noise computation. Moreover, ${\dot L}^2$ terms (and higher order) can be dropped since they are of the order of $\lsim 10^{-15}$ (they come with a factor $1/c^2$) which is much smaller than 1 part in $10^8$. The calculations which follow neglect these terms which simplifies the computations and makes them tractable. We begin with the effect of one operator on $C(t)$ and then by induction obtain the effect $n$ succesive operators operating on $C(t)$.   
\be
D_k C(t) = C(t - L_k(t)) \equiv E_k C(t) \,, 
\ee
where $E_k$ is a delay operator with constant delay by the time $L_k(t)$ at the given time $t$.

Applying the operators twice in succession and dropping higher order terms as explained above,
\bea
D_{k_2} D_{k_1} C &=& C(t - L_{k_1} (t - L_{k_2}) - L_{k_2}) \,, \no \\
    &\approx& E_{k_2}E_{k_1} C + L_{k_2} {\dot L}_{k_1} E_{k_2}E_{k_1} {\dot C} \,.
\eea
It is easy to generalise the above formula by induction to $n$ operators:
\bea
D_{k_n} ... D_{k_1} C &=& E_{k_n} ... E_{k_1} C + f_n E_{k_n} ... E_{k_1} {\dot C} \,, \no \\
f_n &=& \sum_{p=2}^{n} L_{k_{p}} \sum_{q=1}^{p-1} {\dot L}_{k_q} \,.
\eea
It must be noted that both $C$ and its time derivative are evaluated at the delayed time given by the successive application of $n$ delay operators $E_{k_m}$.
\par
We now turn to the commutators of the operators. These occur in many of the LISA observables and therefore it is useful compute these. The term in $C$ cancels out; only the ${\dot C}$ term remains. The simplest of the commutators is:
\be
[D_j, D_k] = D_j D_k - D_k D_j =  L_j {\dot L}_k - L_k {\dot L}_j\,, 
\ee
where it is understood that the commutator multiplies ${\dot C}$ at the delayed time 
$t - L_j(t) - L_k(t)$ for fixed time $t$.
\par
For short we will write $j$ instead of $D_j$ whenever there is no possibility of confusion. Thus the commutator $[D_j, D_k]$ will be simply written by $[j,k]$. We list few more commutators that occur in the observables:
\bea
kmj - jkm &=& (L_k + L_m) \Ldot_j - L_j (\Ldot_k + \Ldot_m) \,, \no \\
lmjk - jklm &=& (L_l + L_m)(\Ldot_j + \Ldot_k) - (L_j + L_k)(\Ldot_l + \Ldot_m) \,, \no \\
lmnxyz - xyzlmn &=& (L_l + L_m + L_n)(\Ldot_x + \Ldot_y + \Ldot_z) \\
                && - (L_x + L_y + L_z)(\Ldot_l + \Ldot_m + \Ldot_n) \,.
\label{commut}
\eea
This formula generalises in an obvious way to $2n$ operators.

\subsection{Residual laser frequency noise in some important TDI observables} 

The laser frequency noise is usually assumed to be $\sim 30 {\rm Hz}/\sqrt{\rm Hz}$ in most of the literature so far. However, by the time LISA flies the expectations are for this noise estimate to reduce to say $\Delta \nu \sim 10 {\rm Hz}/\sqrt{\rm Hz}$. If we divide this number by the laser frequency $\nu_0 \sim 3 \times 10^{14}$ Hz, we obtain the noise estimate in the fractional Doppler shift $C(t) = \Delta \nu (t)/ \nu_0 \sim 3 \times 10^{-14}$. Thus the power spectral density (PSD) of the noise $C$ is:
\be
S_C (f) = \langle |{\tilde C} (f)|^2 \rangle \sim 10^{-27}~~~ {\rm Hz}^{-1} \,,
\label{lsr_noise}
\ee
where ${\tilde C} (f)$ is the Fourier transform of $C(t)$.
\par
From this equation it is easily deduced on differentiating that the PSD of the random variable 
${\dot C}$ is:
\be
S_{\dot C} (f) = 4 \pi^2 f^2 S_C (f)~~~ {\rm Hz} \,.
\ee
The commutators computed in the last subsection when divided by $c^2$ have dimension of time (assuming that the dot on $L_k$ is just $d/dt$). A commutator is essentially the time difference in the transit times of photons along two different paths - the residual time. This is the reason why the laser frequency noise does not cancel out. If the two paths were exactly equal, the laser frequency noise would completely cancel out as it happens when the armlengths are constant. In order to compare the noise in a specific observable, say, the Sagnac or Michelson, one must compare the PSD of the secondary noise in that observable (which has dimensions of ${\rm Hz}^{-1}$) with the residual laser frequency noise PSD which is given by  $\Delta t^2 S_{\dot C}(f) $, where $\Delta t$ is the time difference. Note that $\Delta t$ is a function of $t$ as the LISA constellation evolves and propagates in the gravitational field. We carry out these computations for the well known observables, such as the Sagnac, denoted by 
$\a, \beta, \gamma$, the Michelson observable $X$, the symmetric Sagnac observable $\zeta$. For the observable $\zeta$, the commutators have different degree polynomials and therefore, the ${\dot C}$ appears at different delays. In the Fourier domain, these then appear as phase factors in $\Delta t$, which in effect becomes complex.
\par
We follow the notation and conventions of \cite{NV} and \cite{DNV} which are the simplest for our purpose. The six links are denoted by $U^i, V^i , i = 1, 2, 3$. The time-delay for the link $U^2$ from S/C 1 to S/C 2 or $1 \longrightarrow 2$ is denoted by $x$ in \cite{NV} (which is $3'$ in \cite{Vallis} and so on in a cyclic fashion); the delay for link $U^3$ from $ 2 \longrightarrow 3$ by $y$; the delay for link $U^1$ from $3 \longrightarrow 1$ by $z$. The delays in the other sense are denoted by $l, m, n$. The delay for the link $-V^1$ from $2 \longrightarrow 1$ by $l$; and then links $V^2, V^3$ and the corresponding delays $m, n$ are defined through cyclic permutation. In the formalism any observable $X$ is given by:
\be   
X = p_i V^i + q_i U^i \,,
\ee
where $p_i, q_i, i = 1,2,3$ are polynomial vectors in the variables $x,y,z,l,m,n$. Thus $X$ is specified by giving the six tuple polynomial vector $(p_i, q_i)$.  
\par
We observe the following approximate symmetries in our model:
\be
\Ldot_x \approx \Ldot_l,~~ \Ldot_y \approx \Ldot_m, ~~ \Ldot_z \approx \Ldot_n \,,
\label{comm}
\ee
which also implies (this combination occurs in the Sagnac observables),
\be
\Ldot_x + \Ldot_y + \Ldot_z \approx \Ldot_l + \Ldot_m + \Ldot_n \,.
\label{cyclic}
\ee
It was shown in \cite{NKDV} that only in the Sun's field ${\dot L_{ij}} \propto (\sin \Omega (t-t_{ij}^{(0)}) + k \sin 3 \Omega (t-t_{ij}^{(0)}))$ where $k$ is a constant and $t_{ij}^{(0)}$ are given constants. When we consider the sum $\Ldot_x + \Ldot_y + \Ldot_z$, the phases for the links $y$ and $z$ namely, $\Omega t_{y}^{(0)}$ and $ \Omega t_{z}^{(0)}$ differ by $ 2 \pi / 3$ and $4 \pi / 3$ from the phase of the link $x$ and therefore their sum is close to zero. The same is true for the links $l,m,n$. Here, when we consider the combined field of the Sun and Earth, this is no longer true but we find that,  
$|(\Ldot_x + \Ldot_y + \Ldot_z) - (\Ldot_l + \Ldot_m + \Ldot_n)| \lsim 1$ m/sec and $|\Ldot_x - \Ldot_l| \lsim 0.8$ m/sec upto the first three years in our model. The same is essentially true for the pairs of links $y, m$ and $z, n$. Thus these pairs of operators essentially commute. 
\par
Thus here, for practical purposes, we are not dealing with a set of totally noncommuting variables, but with an intermediate case in which the six variables partition pairwise into three pairs, such that for  each pair the variables essentially commute, while any other combination of variables does not. 

\subsubsection{The Sagnac variables:}

The modified first generation TDI observable $\alpha$ is given by the polynomial vector in the form $(p_i, q_i)$ by:
\be
\alpha = (\kappa, \kappa l, \kappa lm, \eta, \eta zy, \eta z) \,,
\label{sgnc}
\ee 
where $\kappa = 1 - zyx$ and $\eta = 1 - lmn$. If the variables $x,y,z,l,m,n$ commute then the laser frequency noise is fully cancelled. However, if they do not commute, there is a residual term. Let the laser frequency noises on each spacecraft $i$ be $C_i$ respectively (we consider a single effective laser frequency noise random variable on each spacecraft), then the residual term is:
\be
\Delta C = \a_1 C_1 + \a_2 C_2 + \a_3 C_3 \,.
\ee 
We find that $\a_2 = \a_3 \equiv 0$ and $\a_1 = [zyx, lmn]$ and so by Eq. (\ref{commut}): 
\be
\Delta t (t) = \frac{1}{c^2}[(L_x + L_y + L_z)(\Ldot_l + \Ldot_m + \Ldot_n) - (L_l + L_m + L_n)(\Ldot_x + \Ldot_y + \Ldot_z) ] \,,
\ee
and thus $\Delta C = \Delta t {\dot C_1}$. Because the $L_k$ vary during the course of an year the $\Delta t$ also varies during the year and so also the amplitude of the random variable $\Delta C$. Thus the PSD of $\Delta C$ is:
\be
S_{\Delta C} (f;t) = 4 \pi^2 \Delta t(t)^2 f^2 S_C(f) \,.
\label{rsdl}
\ee
This is the residual laser frequency noise in the observable $\alpha$ which depends on the epoch $t$. 
\par
\begin{figure}[h]  
\centering  
\includegraphics[width = 0.6\textwidth]{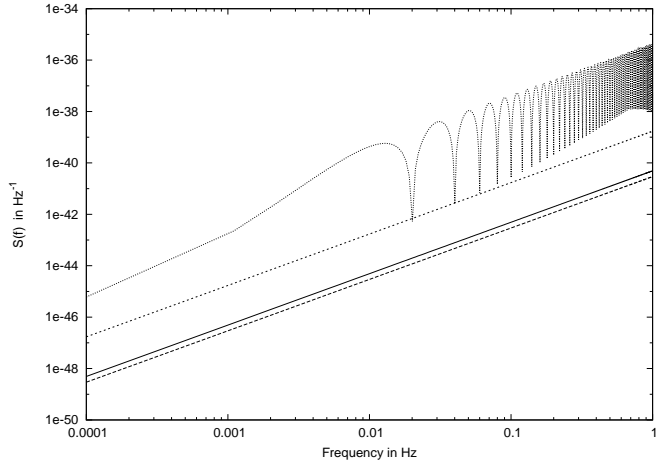}  
\caption{The `top' curve shows the PSD $S_{\a} (f)$ of the secondary noises. The straight lines are the PSDs of the residual noise at three epochs chosen an year apart. Clearly the residual laser noise is adequately below the secondary noises.}  
\label{sagnac}  
\end{figure}

This noise must be compared with the secondary noise \cite{RIP}. However, because we are considering the modified TDI Eq. (\ref{sgnc}), there are extra factors $\kappa$ and $\eta$ which do not appear in the corresponding first generation TDI. These factors introduce an additional multiplicative factor, namely, $4 \sin^2 (3 \pi f L_0)$ in the secondary noise PSD which leaves the SNR unchanged but must be considered when it is compared with the residual laser frequency noise given in Eq. (\ref{rsdl}). Thus,
\be
S_{\a} (f) = 4 \sin^2 (3 \pi f L_0)\{[8 \sin^2 3 \pi f L_0 + 16 \sin^2 \pi f L_0] S_{acc} + 6 S_{opt}\}
\ee
where $S_{acc} = 2.5 \times 10^{-48} (f/1 {\rm Hz})^{-2} {\rm Hz}^{-1}$ and 
$S_{opt} = 1.8 \times 10^{-37} (f/1 {\rm Hz})^2 {\rm Hz}^{-1}$.
In the Figure \ref{sagnac} we plot $S_{\a} (f)$ and $S_{\Delta C} (f;t)$ at three epochs an year apart. 

We see that, clearly the residual laser frequency noise is an order or few orders of magnitude below the secondary noises depending on the epoch. Since the other Sagnac variables $\beta$ and $\gamma$ are obtained by cyclic permutations of the spacecraft, the residual laser noise is the same, if we assume that the noise PSDs of $C_i$ are identical; the $C_1$ is replaced by $C_2$ and $C_3$ respectively for $\beta$ and $\gamma$. Thus modified first generation TDI Sagnac variables suffice to suppress the laser frequency noise.
\par
As noted before, the basic reason for this remarkable cancellation is the symmetry inherent in the spacecraft configuration and physics as detailed in the equations Eq. (\ref{comm}) and Eq. (\ref{cyclic}). 

\subsubsection{The Michelson variables:}

\begin{figure}[t]  
\centering  
\includegraphics[width = 0.6\textwidth]{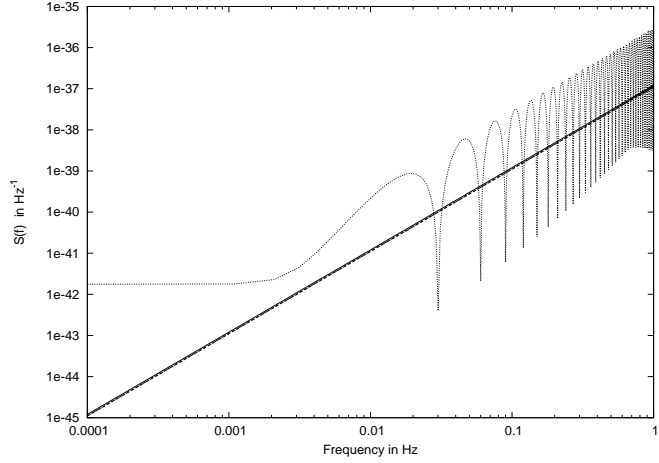}  
\caption{The curve shows the PSD $S_{X} (f)$ of the secondary noises. The straight lines are the PSDs of the residual noise at three epochs an year apart. The residual noises are essentially below the secondary noises. Also we note that in this case the straight lines lie very close to each other suggesting  periodicity of a year.}  
\label{mich}  
\end{figure}

The TDI observable $X$ is given by:
\be
X = (1-zn, 0, (lx - 1)z, 1 - lx, (zn - 1)l, 0)
\ee
The residual term is:
\be
\Delta C = X_1 C_1 + X_2 C_2 + X_3 C_3 \,.
\ee 
It is found that $X_2 = X_3 \equiv 0$ and $X_1 = [zn, lx]$ and hence:
\be
\Delta t (t) = \frac{1}{c^2}[(L_z + L_n)(\Ldot_l + \Ldot_x) - (L_l + L_x)(\Ldot_z + \Ldot_n) ]
\ee
Using this value of $\Delta t$ in Eq. (\ref{rsdl}) and Eq. (\ref{lsr_noise}) gives the residual noise in the Michelson variable $X$. The PSD of the secondary noise is given by \cite{RIP}:
\be
S_X (f) = [8 \sin^2 4 \pi f L_0 + 32 \sin^2 2 \pi f L_0] S_{acc} + 16 \sin^2 2 \pi f L_0 S_{opt}
\ee
The noise plots are shown in Figure \ref{mich}. Clearly at low frequencies $f \lsim 1$ mHz the TDI variable $X$ suffices, that is, the laser frequency noise is adequately suppressed. At higher frequencies in most of the frequency domain the residual noise remains below 20$\%$.

\subsubsection{The symmetric Sagnac variables:}
 
The modified symmetric sagnac variables denoted by $\zeta$ in the literature split into three. We consider here just one of these because the noise in the others is essentially the same. It is given by:
\be
\zeta = (y(zx-m), (ln-y)z, (zx-m)l, m(ln-y), (ln-y)z, (zx-m)l) \,.
\ee
We denote it by just $\zeta$, dropping the subscript, instead of $\zeta_1$, because we will not be explicitly discussing the other two cyclic permutations of $\zeta$. The residual laser frequency noise term can again be written as $\Delta C = \zeta_1 C_1 + \zeta_2 C_2 + \zeta_3 C_3$ where the $\zeta_k$ can be expressed in terms of the commutators,
\bea
\zeta_1 &=& [m,y] + [ln, zx] \, \no \\
\zeta_2 &=& [(zx-m)l,y] \, \no \\
\zeta_3 &=& [m, (ln-y)z] \,.
\eea
Note that here none of the $\zeta_k$ are zero and hence contribute to the total residual noise. Here it is more appropriate to compute the random variables ${\Delta C}_k$ defined below in terms of the commutators given above and the delay operators $E_k$:
\bea
c^2 {\Delta C}_1 &=& [- L_y \Ldot_m + L_m \Ldot_y]E_m E_y {\dot C}_1 \, \no \\ 
                 && + [(L_n + L_l)(\Ldot_x + \Ldot_z) - (L_x + L_z)(\Ldot_n + \Ldot_l)]E_x E_z E_n E_l {\dot C}_1  \, \no \\
c^2 {\Delta C}_2 &=& [L_y (\Ldot_l + \Ldot_m) - \Ldot_y (L_l + L_m)]E_m E_l E_y {\dot C}_2  \no \\
               && + [\Ldot_y(L_x + L_z + L_l) - L_y(\Ldot_x + \Ldot_z + \Ldot_l)]E_y E_l E_x E_z {\dot C}_2 \,, \no \\
c^2 {\Delta C}_3 &=& [- L_m (\Ldot_y + \Ldot_z) + \Ldot_m (L_y + L_z)]E_m E_y E_z {\dot C}_3  \no \\
               && + [- \Ldot_m(L_n + L_z + L_l) + L_m(\Ldot_n + \Ldot_z + \Ldot_l)]E_m E_l E_n E_z {\dot C}_3 \,\,.
\eea
We note here that in each equation, the ${\dot C}_k$ is delayed by different amounts. In the Fourier space this is translated into complex phase factors. For the purpose of this computation assuming equal armlengths $L_0$ for all the links, the effective complex $\Delta t_1$ is given by:
\bea
c^2 \Delta t_1 (t) &=& [- L_y \Ldot_m + L_m \Ldot_y]e^{-4 \pi i f L_0} \no \\ 
&& + [(L_n + L_l)(\Ldot_x + \Ldot_z) - (L_x + L_z)(\Ldot_n + \Ldot_l)] e^{- 8 \pi i f L_0} \,. 
\eea
Similar expressions can be derived for $\Delta t_2$ and $\Delta t_3$. When computing the PSDs it is the modulus of $\Delta t_k$ that enters into their expressions accounting for the phases. 
\par
\begin{figure}[t]  
\centering  
\includegraphics[width = 0.6\textwidth]{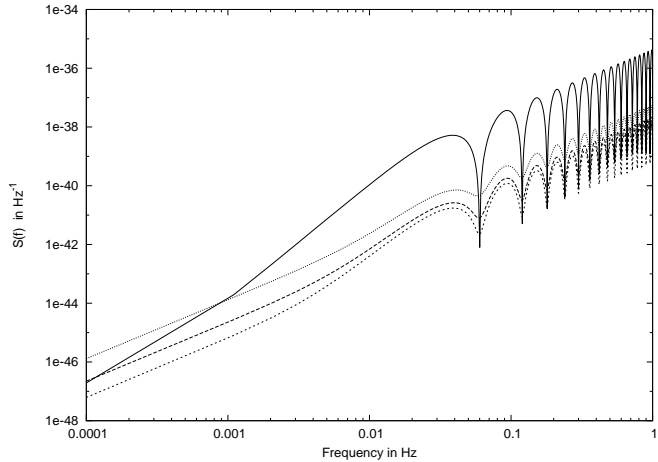}  
\caption{The bold curve shows the PSD $S_{\zeta} (f)$ of the secondary noises. The rest of the curves  are the PSDs of the residual noise at three epochs an year apart. The residual noises are below the secondary noises except near the low frequency end.}  
\label{zeta}  
\end{figure}

We now assume that laser noises $C_k$ are independent of each other and also have identical PSDs. Therefore we may add the noises quadratically - that is, we take the sum of the PSDs. The result is:
\be
S_{\Delta C} (f;t) = 4 \pi^2 f^2 (|\Delta t_1|^2 + |\Delta t_2|^2 + |\Delta t_3|^2) S_C(f) \,.
\ee
This equation describes the residual laser noise. We must now compare this PSD with the sum of the PSDs of  optical and acceleration noises. This PSD is given by:
\be
S_{\zeta} (f) = 4 \sin^2 \pi f L_0 (24 \sin^2 \pi f L_0 S_{acc} + 6 S_{opt}) \,.
\ee 
There is an extra factor of $4 \sin^2 \pi f L_0$ in the modified TDI variable zeta in the PSD which must be considered when comparing with the residual laser frequency noise. The noise PSDs are plotted in Figure \ref{zeta}. 
At the low frequency end $f \lsim 1$ mHz, the residual noise is close to $S_{\zeta} (f)$, while at higher frequencies it less by an order of magnitude and thus reasonably suppressed. 

\section{Concluding remarks}

 In this work we have included the effect of the Earth on the flexing of the arms of LISA. We have computed the spacecraft orbits in the combined field of the Sun and Earth approximately and from this deduced the flexing of the arms of LISA by choosing the model which gave minimum flexing when only the Sun's field was taken into account. We note that the flexing in the combined field is no more periodic as was the case when only the Sun's field was considered. We have ignored the effect of Jupiter because we believe this effect to be not so dominant as that of the Earth. Firstly, if we compute the tidal parameter $\eps_J$ for Jupiter, $\eps_J = G M_J/d_J^3$, similar to $\eps$ of Earth, where, $M_J \approx 2 \times 10^{27}$ kg is the mass of Jupiter and $d_J$, the distance from LISA to Jupiter, which we take on the average to be $\sim 5$ A. U., then $\eps_J/\eps \sim 0.09$. Thus we expect the effect of Jupiter to be less than 10$\%$ than that due to Earth. Secondly, the forcing terms of Jupiter have the periodicity pertaining to its own orbit and therefore will not be in resonance as was the case with the Earth, so we do not expect the effect to accumulate in the first few years, when the perturbations are small. Note that these results are valid so long as we can neglect the nonlinearities arising from higher order terms in $\eps$ and $\a$.  
\par
We have then used the results of the flexing of LISA's arms to compute the residual laser frequency noise in important TDI variables, namely, the Sagnac, the Michelson and the Symmetric Sagnac. Our results are obtained to the first order in $\Ldot$ dropping terms of degree/order equal to or higher than $\Ldot^2$ and ${\ddot L}$. We have compared the residual noises with the corresponding secondary noises. We find that the residual laser noise in all these  variables tends not to be very high. In the Sagnac variables it is negligible, in the Michelson variables it is less than 20$\%$, while in $\zeta$ variables only at the low frequency end the residual noise becomes comparable to the secondary noises. If this is acceptable then the modified first generation TDI observables could as well be used along with our model of LISA. Second generation TDI variables generally involving higher degree polynomials may not be then required.
\par
Our model of LISA is optimal (minimal flexing of arms) only in the Sun's field. Clearly this opens up the question of seeking an optimal model for the LISA configuration in the field of the Sun, Earth, Jupiter and other planets which will minimise the flexing of the arms and therefore the residual laser frequency noise in the modified first generation TDI. 
\par
We finally remark that our computations here may be useful in the development of a LISA simulator. This is because our computation of the optical links have been carried out within a fully general relativistic framework and we have taken into account the gravitational field of the Earth. Also any discrepancy observed between actual data and the model may suggest physical causes which would be of interest and therefore would have to be incorporated into the data analysis.
\ack

The authors would like to thank the Indo-French Centre for the Promotion of Advanced Research (IFCPAR) project no. 3504-1 under which this work has been carried out. RN would like to acknowledge IUCAA for local hospitality, where part of this work was carried out.


\vspace{24pt}   


\begin{thebibliography}{10}
  
\bibitem{GWD} A.~Abramovici {\it et al.}, Science {\bf 256}, 325 (1992); C.~Bradaschia {\it et al.}, Nucl.\ Instrum.\ Methods\ Phys.\ Res.\ A {\bf 289}, 518, (1990); K.~Danzmann, in {\em Gravitational Wave Experiments}, edited by E.~Coccia, G.~Pizzella and F.~Ronga (World Scientific, Singapore, 1995), pp. 100--111;  K.~Tsubono, in {\em Gravitational Wave Experiments}, edited by E.~Coccia, G.~Pizzella and F.~Ronga (World Scientific, Singapore, 1995), pp. 112--114; R.~J.~Sandeman {\it et al.}, A.I.G.O. Prospectus (1997), unpublished.
\bibitem{RIP} P. Bender \emph{et al.} \char`\"{}LISA: A Cornerstone Mission for
the Observation of Gravitational Waves\char`\"{}, System and Technology
Study Report ESA-SCI(2000) 11, 2000. 
\bibitem{Arm} J. W. Armstrong, lrr-2006-1 : http://relativity.livingreviews.org/Articles/lrr-2008-2.
\bibitem{DNV} S. V. Dhurandhar, K. Rajesh Nayak, J-Y. Vinet, \textit{Phys. Rev},
\textbf{D 65} , 102002(2002).
\bibitem{CW} W. H. Clohessy and R. S. Wiltshire, Journal of Aerospace Sciences,  653 - 658 (1960); \\ 
D. A. Vallado, {\it Foundations of Astrodynamics and
Applications}, 2nd edition 2001, Microcosm Press Kluwer;  \\ 
also in S. Nerem, http://ccar.colorado.edu/asen5050/lecture12.pdf(2003). 
\bibitem{LISACode} A. Petiteau, G. Auger, H. Halloin, O. Jeannin, E. Pagnol, S. Pireaux, T. Regimbau and J-Y. Vinet, {\it Phys. Rev.} D {\bf 77}, 023002 (2008).
\bibitem{NKDV} S. V. Dhurandhar, K. R. Nayak, S. Koshti and J-Y. Vinet, {\it Class. Quantum Grav.}, {\bf 22}, 481 (2005); R. Nayak, S. Koshti, S. V. Dhurandhar and J-Y. Vinet, {\it Class. Quantum Grav.}, {\bf 22}, 1763 (2006).
\bibitem{NV} K. R. Nayak and J-Y Vinet, {\it Phys. Rev.} D {\bf 70}, 102003 (2004).  
\bibitem{Vallis} M. Vallisneri, {\it Phys. Rev.} D {\bf 72}, 04003 (2005).  

\end{thebibliography}
\end{document}